\documentclass{article}
\usepackage{geometry}
\usepackage{enumerate}
\usepackage{amssymb,amsfonts,amsmath,bbm,mathrsfs,stmaryrd,mathtools,mathabx}
\usepackage{amsmath}
\usepackage{amsfonts,amscd, amssymb}
\usepackage{xcolor}
\usepackage{tikz-cd}
\usepackage{amsthm,latexsym,euscript,amscd, braket}
\usepackage{aliascnt}
\usepackage{amsmath,amsthm}
\usepackage{thmtools, ulem}
\usepackage{xcolor,colortbl}
\definecolor{lavender}{rgb}{0.9, 0.9, 0.98}
\definecolor{UGARed}{rgb}{0.729,0.047,0.184}
\usepackage{url}
\usepackage{authblk}
\usepackage{enumerate}
\usepackage{thmtools}
\usepackage[backref=page]{hyperref}
\newcommand\nnfootnote[1]{%
  \begin{NoHyper}
  \renewcommand\thefootnote{}\footnote{#1}%
  \addtocounter{footnote}{-1}%
  \end{NoHyper}
}
\usepackage{cleveref}
\creflabelformat{enumi}{#2(#1)#3}
\crefformat{enumi}{(#2#1#3)}
\crefformat{equation}{(#2#1#3)} 
\Crefformat{equation}{(#2#1#3)}

\crefformat{inequality}{(#2#1#3)} 
\Crefformat{Inequality}{(#2#1#3)}
\hypersetup{colorlinks,
linkcolor=UGARed,
citecolor=blue,
pdftitle={},
bookmarksopen=true,
bookmarksnumbered=true}         
\usepackage{tikz}
\usepackage{graphicx} 
 \usepackage{todonotes} 
\theoremstyle{plain} 
\newtheorem{thm}{Theorem}[section]
\newtheorem{cor}[thm]{Corollary}

\newtheorem{rmk}[thm]{Remark}

\numberwithin{equation}{section}

\DeclareMathOperator{\Al}{{\mathcal A}}

\newcommand{\norm}[1]{\left\Vert#1\right\Vert}

\begin{document}
\title{Suzuki Type Estimates for Exponentiated Sums 	
	and Generalized Lie-Trotter Formulas  
in Banach Algebras}


\author{Zhenhua Wang \thanks{\texttt{zhenhua.wang@aamu.edu}}}

\affil{Department of Physics, Chemistry and Mathematics, Alabama A$\&$M University}

\date{\today}
\maketitle

\begin{abstract}
  The Lie-Trotter formula has been a fundamental tool in quantum mechanics, quantum computing, and quantum simulations. The error estimations for the Lie-Trotter product formula play a crucial role in achieving scalability and computational efficiency. In this note, we present two error estimates of Lie-Trotter product formulas, utilizing Jordan product within Banach algebras. Additionally, we introduce two generalized Lie-Trotter formula and provide two explicit estimation formulas. Consequently, the renowned Suzuki symmetrized approximation for the exponentiated sums follows directly from our main Theorem.
\end{abstract}

\nnfootnote{\textit{Mathematics Subject Classification \rm{(2020)}:} {47A58, 
	47N50,
	17C65;
	Secondary 
	15A16, 
	81R15,
	81P45
	}}

\nnfootnote{\textit{Key words:} Lie-Trotter formula, Suzuki approximation, Banach algebra, Jordan product}
\vspace{0.25cm}

\section{Introduction}
The Trotter product formula \cite{trotter:1959:product} has played a pivotal role in the fields of quantum mechanics, quantum computing, and quantum simulationsdue to its capability to decompose complex quantum systems into simpler subsystems. Accurate estimation of errors associated with the Lie-Trotter product formula is crucial for achieving scalability and computational efficiency in various quantum applications.

The theory of Trotter product formulas in Banach algebras was significantly advanced by Suzuki. In 1976, Suzuki introduced estimates for the product formula approximation, extending the Lie-Trotter formula to exponentiated sums of an arbitrary number of elements in a Banach algebra \cite{suzuki:1976:generalized}. He later refined these estimates in 1985 by presenting a symmetrized approximation method for exponentiated sums \cite{suzuki:1985:transfer}. The applications of his results are extensive, notably in quantum chemistry, condensed matter physics, and quantum computing \cite{Poulin:etal:chemistry}.

The extension of Trotter product formulas to Jordan-Banach algebras was first achieved by Peralta and collaborators \cite{escolano:peralta:villena:2024:lie}. 
Inspired by their work, we recently developed Trotter product formulas  in JB-algebras, adopting a distinct approach from Peralta et al. Remarkably, Suzuki-type estimates remain valid even in the non-associative context of JB-algebras \cite{chehade:wang:wang:2024:suzuki}. This theory has further implications for quantum information science \cite{wang:chehade:dumitrescu:2024:semi}. Building on our results \cite{chehade:wang:wang:2024:suzuki, wang:wang:2021:means, wang:wang:2022:refined, wang:wang:2022:entropy}, we introduced the concept of Lie-Trotter means within JB-algebras \cite{wang:2023:lie}.

In this note, we present two error estimates for Lie-Trotter product formulas using Jordan product within Banach algebras. Additionally, we establish two generalized Lie-Trotter formulas and explicitly provide two estimation formulas with an order of $\displaystyle 1/n^2$. Consequently, the notable Suzuki estimate for symmetrized Lie-Trotter formula \cite[Formula 3, Equation (1.15)]{suzuki:1985:transfer} follows directly from our results.  Note that our main results, \Cref{thm:suzuki:1order} and \Cref{thm:suzuki:nonsy},  partially answer the open problem regarding the existence of Trotter error estimates for Jordan-Trotter product formulas in Jordan-Banach algebras, initially posed in \cite{chehade:wang:wang:2024:suzuki}, since every Banach algebra is also a Jordan-Banach algebra. This open problem has recently been fully solved in \cite{chehade:delgado:wang:wang:2024:error}. 

We now introduce necessary notation. Let $\Al$ be a unital Banach algebra. It possesses an underlying Jordan algebra structure with the Jordan product defined by 
\[
A\circ B=(AB+BA)/{2},\, \mbox{for any}\, A, B \in \Al.	
\]
 Note that $A\circ B=B\circ A$ and $\norm{A\circ B}\leq \norm{A}\cdot\norm{B}.$
   
The Jordan triple product on $\Al$ is defined by
\[\displaystyle \{ABC\}:=(A\circ B)\circ C+(C\circ B)\circ A- (A\circ C)\circ B=(ABC+CBA)/2.\] 
It follows readily that $\displaystyle \norm{\{ABC\}}\leq \norm{A}\norm{B}\norm{C}.$ For further details on Jordan products, we refer readers to, we refer the reader to \cite{alfsen:2003:geometry, blecher:wang:2018,blecher:wang:2019, cabrera:rodriguez:2014:non, hanche:1984:jordan}.

\section{Error Estimations for Second Order Jordan-Trotter Product Formulas}\label{section: error}
In this section, we utilize the Jordan product structure within Banach algebras to establish Suzuki-type estimations for Trotter product formulas and explore their implications.

The following theorem provides Suzuki-type estimations for exponentiated sums using Jordan product decompositions in Banach algebras, with an error order of $1/n^2$.
\begin{thm}\label{thm:suzuki:1order}
	For any finite number of elements $\displaystyle A_1, A_2,\cdots, A_m$ in a Banach algebra $\Al,$
	\begin{align*}
		\left\Vert \exp \left( \sum_{j=1}^m A_j\right)-g_{n}(\{A_j\})\right \Vert	
		\leq
		\dfrac{1}{3n^2}\left( \sum_{j=1}^m \Vert A_j\Vert\right)^3\exp\left(\sum_{j=1}^m \Vert A_j\Vert\right),	
	\end{align*}
where
\begin{align*}
g_{n}\left(\{A_j\}\right)&=\left(\left[\left(\exp\left(\frac{A_1}{n}\right)\circ \exp\left(\frac{A_2}{n}\right)\right)\circ\cdots \right]\circ \exp\left(\frac{A_m}{n}\right)\right)^n	
\end{align*}	
\end{thm}
\begin{proof}
	Let
	\begin{align}
		C&=\exp\left(\frac{1}{n}\sum_{j=1}^m A_j\right),\label{NotionC}\\
		D&=\left[\left(\exp\left(\frac{A_1}{n}\right)\circ \exp\left(\frac{A_2}{n}\right)\right)\circ\cdots \right]\circ \exp\left(\frac{A_m}{n}\right)\nonumber.
	\end{align}
	Then,
	\begin{align*}
		\Vert C\Vert\leq\exp \left(\frac{1}{n}\sum_{j=1}^m \Vert A_j\Vert\right)\,\ \mbox{and}\,\  
		\Vert D\Vert\leq \exp \left(\frac{1}{n}\sum_{j=1}^m \Vert A_j\Vert\right).
	\end{align*}
	
We observe that
	\begin{align}
		s&=\left\Vert \exp \left( \sum_{j=1}^m A_j\right)-g_{n}(\{A_j\})\right \Vert	=\left\Vert C^n-D^n \right\Vert \nonumber \\
		&\leq \Vert C-D\Vert\left(\Vert C\Vert^{n-1}+\Vert C\Vert^{n-2}\Vert D\Vert+\cdots+\Vert D\Vert^{n-1}\right)\nonumber \\
		&\leq  n \Vert C-D\Vert\left(\max\{\Vert C\Vert, \Vert D\Vert\}  \right)^{n-1} \nonumber \\
		&\leq n \Vert C-D\Vert \exp \left(\frac{n-1}{n}\sum_{j=1}^m \Vert A_j\Vert\right)\nonumber \\
		&\leq n \left(\Vert C-F\Vert +\Vert D-F\Vert\right)\exp \left(\frac{n-1}{n}\sum_{j=1}^m \Vert A_j\Vert\right), \label{GHnbound3} 
	\end{align}
	where
	\begin{align}
		F=	I+\left( \sum_{j=1}^m \frac{A_j}{n}\right)+\left( \sum\limits_{j=1}^m A_j/n \right)^2/{2!}. \label{NotionF}
	\end{align}
	Note that 
	\begin{align}
		\Vert C-F\Vert&=\left\Vert \sum_{k=3}^{\infty}\left(\frac{A_1+A_2\cdots+A_m}{n}\right)^k\frac{1}{k!}\right\Vert \nonumber \\
		&\leq \exp \left(\frac{1}{n}\sum_{j=1}^m \Vert A_j\Vert \right)-\left(I+\dfrac{\sum_{j=1}^m \Vert A_j\Vert}{n}+\dfrac{\left(\sum_{j=1}^m \Vert A_j\Vert\right)^2}{2n^2}\right) \nonumber \\\
		&\leq \frac{1}{3!\cdot n^3}\left(\sum_{j=1}^m \Vert A_j\Vert \right)^3\exp \left(\frac{1}{n}\sum_{j=1}^m \Vert A_j\Vert \right) \label{TaylorC} 
	\end{align}
	
where the inequality \Cref{TaylorC} follows from \cite[Theorem 1]{suzuki:1976:generalized}.

Following a similar argument as in the proof of Theorem 3.1 in \cite{chehade:wang:wang:2024:suzuki}, the degree 2 Taylor polynomial of $D$ is $F.$ Thus,
	\begin{align}
		\Vert D-F\Vert&= \left\Vert \sum_{k=3}^{\infty}\left(\mbox{sum of all terms of degree $k$ of Taylor expansion of}\, D\right) \right\Vert \nonumber \\ 
  &=\left\Vert \sum_{k_1+\cdots+k_m=3}^{\infty} \dfrac{\left[(A_1^{k_1}\circ A_2^{k_2})\circ \cdots\right]\circ A_m^{k_m} }{n^{k_1+k_2+\cdots+k_m}\cdot k_1!\cdot k_2!\cdots k_m!}\right\Vert \nonumber \\ 
		&\leq \sum_{k_1+\cdots+k_m=3}^{\infty} \dfrac{\Vert A_1\Vert^{k_1} \Vert A_2\Vert^{k_2}\cdots \Vert A_m \Vert ^{k_m} }{n^{k_1+k_2+\cdots+k_m}\cdot k_1!\cdot k_2!\cdots k_m!} \nonumber \\ 
		&= \exp \left(\frac{1}{n}\sum_{j=1}^m \Vert A_j\Vert \right)-\left(I+\dfrac{\sum_{j=1}^m \Vert A_j\Vert}{n}+\dfrac{\left(\sum_{j=1}^m \Vert A_j\Vert\right)^2}{2n^2}\right) \nonumber \\ 
		&\leq \frac{1}{3!\cdot n^3}\left(\sum_{j=1}^m \Vert A_j\Vert \right)^3\exp \left(\frac{1}{n}\sum_{j=1}^m \Vert A_j\Vert \right) \label{TaylorD}
	\end{align}
	where the inequality (\ref{TaylorD}) follow from \cite[Theorem 1]{suzuki:1976:generalized}.
	
	Combining \Cref{GHnbound3}, \Cref{TaylorC} and \Cref{TaylorD}, we get desired result.    
\end{proof}

As a corollary, we have the following generalized Lie-Trotter formula for an 
arbitrary number of elements in a Banach algebra. 

\begin{cor}
	For any finite number of elements $A_1, A_2,\cdots, A_m$ in a Banach algebra $\Al,$
	\begin{align*}
		\lim\limits_{n\to \infty}\left\{\left[\left(\exp\left(\frac{A_1}{n}\right)\circ \exp\left(\frac{A_2}{n}\right)\right)\circ\cdots \right]\circ \exp\left(\frac{A_m}{n}\right)\right\}^n=\exp\left(\sum_{j=1}^m A_j\right).
	\end{align*}
	If $\displaystyle m=2,$ then it is reduced to the following Lie-Trotter formula:
	\begin{align*}
		\exp(A+B)=\lim\limits_{n\to \infty}\left[\frac{1}{2}\left(\exp\left(\frac{A}{n}\right) \exp\left(\frac{B}{n}\right)+\exp\left(\frac{B}{n}\right) \exp\left(\frac{A}{n}\right)\right)\right]^n	
	\end{align*}
	for any elements $A, B$ in $\Al.$	
\end{cor}
\section{Error Estimations for Jordan-Trotter Triple Product Formulas}\label{section: triple}

In this section, we further extend our analysis by introducing Suzuki-type estimations for exponentiated sums decomposed via the Jordan triple product in Banach algebras.
\begin{thm}\label{thm:suzuki:nonsy}
	For any finite number of elements $A_1, A_2,A_3\cdots, A_{2p}, A_{2p+1}$ in a unital Banach algebra $\Al,$
	\begin{align*}
	\left\Vert \exp \left( \sum_{j=1}^{2p+1} A_j\right)-h_{n}(\{A_j\})\right \Vert	
		\leq \frac{1}{3n^2}\left( \sum_{j=1}^{2p+1} \Vert A_j\Vert\right)^3\exp\left(\sum_{j=1}^{2p+1} \Vert A_j\Vert\right)	
	\end{align*}
where
\begin{align*}
h_{n}(\{A_j\})=\left\{\exp\left(\frac{A_{2p}}{n}\right)\cdots\left\{\exp\left(\frac{A_2}{n}\right)\exp\left(\frac{A_1}{n}\right)\exp\left(\frac{A_3}{n}\right) \right\}\cdots	\exp\left(\frac{A_{2p+1}}{n}\right)\right\}^n.	
\end{align*}	
\end{thm}
\begin{proof}
	Let
	\begin{align*}
		G&=\exp\left(\frac{1}{n}\sum_{j=1}^m A_j\right),\\
		H&=\left\{\exp\left(\frac{A_{2p}}{n}\right)\cdots\left\{\exp\left(\frac{A_2}{n}\right)\exp\left(\frac{A_1}{n}\right)\exp\left(\frac{A_3}{n}\right) \right\}\cdots	\exp\left(\frac{A_{2p+1}}{n}\right)\right\}.
	\end{align*}
	Then
	\begin{align*}
		\Vert G\Vert\leq\exp \left(\frac{1}{n}\sum_{j=1}^{2p+1} \Vert A_j\Vert\right) \quad \mbox{and}\quad  
		\Vert H\Vert \leq \exp \left(\frac{1}{n}\sum_{j=1}^{2p+1} \Vert A_j\Vert\right).
	\end{align*}
	
Note that	
	\begin{align}
		t&=\left\Vert \exp \left( \sum_{j=1}^m A_j\right)-h_{n}(\{A_j\})\right \Vert	=\left\Vert G^n-H^n \right\Vert \nonumber \\
		&\leq\Vert G-H\Vert\left(\Vert G\Vert^{n-1}+\Vert G\Vert^{n-2}\Vert H\Vert+\cdots+\Vert H\Vert^{n-1}\right)\nonumber \\
		&\leq  n \Vert G-H\Vert\left\{\max\{\Vert G\Vert, \Vert H\Vert\}  \right\}^{n-1} \nonumber \\
		&\leq n \Vert G-H\Vert \exp \left(\frac{n-1}{n}\sum_{j=1}^m \Vert A_j\Vert\right) \label{GHnbound}
	\end{align}
	
	For any $\displaystyle 1\leq k\leq p$, let
	\begin{align*}
		E_k=	1+\left(\sum_{j=1}^{2k+1}\frac{A_j}{n}\right)+\dfrac{\left(\sum\limits_{j=1}^{2k+1}\frac{A_j}{n}\right)^2}{2!},
	\end{align*}
	We claim that the degree 2 Taylor polynomials of $\displaystyle G$ and $\displaystyle H$ are $E_{p}$. It is trival for $\displaystyle G$. For any $1\leq k\leq p,$ let
	\begin{align*}
		H_k=\left\{\exp\left(\frac{A_{2k}}{n}\right)\cdots\left\{\exp\left(\frac{A_2}{n}\right)\exp\left(\frac{A_1}{n}\right)\exp\left(\frac{A_3}{n}\right) \right\}\cdots	\exp\left(\frac{A_{2k+1}}{n}\right)\right\}.	
	\end{align*}

	The degree 2 Taylor polynomial of $\displaystyle H_1$ is
	the degree 2 Taylor polynomial of 
	\begin{align*}
		\left(\exp\left(\frac{A_2}{n}\right)\circ \exp\left(\frac{A_1}{n}\right)\right)\circ \exp\left(\frac{A_3}{n}\right)
		+\left(\exp\left(\frac{A_3}{n}\right)\circ \exp\left(\frac{A_1}{n}\right)\right)\circ
		 \exp\left(\frac{A_2}{n}\right)\\
		-\left(\exp\left(\frac{A_2}{n}\right)\circ \exp\left(\frac{A_3}{n}\right)\right)\circ \exp\left(\frac{A_1}{n}\right).	
	\end{align*} 
	In the proof of Theorem 3.1 in \cite{chehade:wang:wang:2024:suzuki}, it is shown that the degree 2 Taylor polynomial of 
	\begin{align*}
	\left(\exp\left(\frac{A_2}{n}\right)\circ \exp\left(\frac{A_1}{n}\right)\right)\circ \exp\left(\frac{A_3}{n}\right)
	\end{align*}
	is $F_1.$ Similarly, it is easy to see that
	the degree 2 Taylor polynomial of 
	\begin{align*}
	\left(\exp\left(\frac{A_3}{n}\right)\circ \exp\left(\frac{A_1}{n}\right)\right)\circ
		 \exp\left(\frac{A_2}{n}\right)
	\end{align*}
 is equal to the degree 2 Taylor polynomial of 
	\begin{align*}
	\left(\exp\left(\frac{A_2}{n}\right)\circ \exp\left(\frac{A_3}{n}\right)\right)\circ \exp\left(\frac{A_1}{n}\right),
	\end{align*}
	which is $F_1.$
Therefore, the degree 2 Taylor polynomial of $\displaystyle H_1$ is $E_1.$

	We observe that the degree 2 Taylor polynomial of $\displaystyle H_2$ is exactly
	the degree 2 Taylor polynomial of 
	\begin{align*}
		\left\{\left(1+\frac{A_4}{n}+\dfrac{\left(\frac{A_4}{n}\right)^2}{2!}\right)E_1\left(1+\frac{A_5}{n}+\dfrac{\left(\frac{A_5}{n}\right)^2}{2!}\right)\right\}.
	\end{align*} 
A direct computation shows that the above is
	\begin{align*}
		E_2=1+\left(\sum_{j=1}^{5}\frac{A_j}{n}\right)+\dfrac{\left(\sum\limits_{j=1}^{5}\frac{A_j}{n}\right)^2}{2!}
	\end{align*}
	By induction, the degree 2 Taylor polynomial of $\displaystyle H$ is $E_p.$ 
	
	By \Cref{TaylorC}, 
	\begin{align}
		\left\Vert G-E_p\right \Vert\leq \frac{1}{3!\cdot n^3}\left(\sum_{j=1}^{2p=1} \Vert A_j\Vert \right)^3\exp \left(\frac{1}{n}\sum_{j=1}^{2p+1} \Vert A_j\Vert \right) \label{TaylorG}
	\end{align}
	since $\displaystyle G=C,$ which is defined \Cref{NotionC}.
	
	Since
	\begin{align*}
		H_1=\frac{1}{2}\left[\exp\left(\frac{A_2}{n}\right)\exp\left(\frac{A_1}{n}\right)\exp\left(\frac{A_3}{n}\right)+\exp\left(\frac{A_3}{n}\right)\exp\left(\frac{A_1}{n}\right)\exp\left(\frac{A_2}{n}\right)\right],	
	\end{align*}
	for any integer $p>2,$ the norm of the sum of all terms of degree $p$ of Taylor expansion of $H_1\leq$ the sum of all terms of degree $p$ of Taylor expansion of \\ 
	$\displaystyle \exp\left(\Vert A_1\Vert/n\right)\exp\left(\Vert A_2\Vert/n\right)\exp\left(\Vert A_3\Vert/n\right).$
	
	Therefore, 
	\begin{align*}
		\left\Vert H_1-E_1\right\Vert&=\left\Vert \sum_{l>2}\left(\mbox{sum of all terms of degree $l$ of Taylor expansion of}\, H_1\right) \right\Vert \\
		&\leq \exp \left(\frac{1}{n}\sum_{j=1}^3 \Vert A_j\Vert \right)-\left(I+\frac{1}{n}\sum_{j=1}^3 \Vert A_j\Vert+\frac{1}{2n^2}\left(\sum_{j=1}^3 \Vert A_j\Vert\right)^2 \right)	
	\end{align*}
	which follows from the fact that the degree 2 Taylor polynomial of $\displaystyle H_1-E_1$ is zero.
		
	Note that 
	\begin{align*}
		H_2=\frac{1}{2}\left[\exp\left(\frac{A_4}{n}\right)H_1\exp\left(\frac{A_5}{n}\right)+\exp\left(\frac{A_5}{n}\right)H_1\exp\left(\frac{A_4}{n}\right)\right].	
	\end{align*}
	For any positive integer $p>2,$ the norm of the sum of all terms of degree $p$ of Taylor expansion of $H_2\leq$ the sum of all terms of degree $p$ of Taylor expansion of $\displaystyle \exp\left(\sum_{j=1}^5\Vert A_j\Vert/n\right).$
	
	Therefore, 
	\begin{align*}
		\left\Vert H_2-E_2\right\Vert&= \left\Vert\sum_{l>2}\left(\mbox{sum of all terms of degree $l$ of Taylor expansion of}\, H_2\right) \right\Vert \\
		&\leq \exp \left(\frac{1}{n}\sum_{j=1}^5 \Vert A_j\Vert \right)-\left(I+\frac{1}{n}\sum_{j=1}^5 \Vert A_j\Vert+\frac{1}{2n^2}\left(\sum_{j=1}^5 \Vert A_j\Vert\right)^2 \right) 	
	\end{align*}
	which follows from the fact that the degree 2 Taylor polynomial of $\displaystyle H_2-E_2$ is zero.
	
	By induction, 
	\begin{align}
		\left\Vert H_p-F_p\right\Vert &=\left\Vert H-F\right \Vert \nonumber \\
		&\leq \exp \left(\frac{1}{n}\sum_{j=1}^{2p+1} \Vert A_j\Vert \right)-\left(I+\frac{1}{n}\sum_{j=1}^{2p+1} \Vert A_j\Vert+\frac{1}{2n^2}\left(\sum_{j=1}^{2p+1} \Vert A_j\Vert\right)^2 \right) \nonumber \\
		&\leq \frac{1}{3!\cdot n^3}\left(\sum_{j=1}^{2p+1} \Vert A_j\Vert \right)^3\exp \left(\frac{1}{n}\sum_{j=1}^{2p+1} \Vert A_j\Vert \right) \label{TaylorH}
	\end{align}
	where the inequality \Cref{TaylorH} follows from \cite[Theorem 1]{suzuki:1976:generalized}.
		
	From \Cref{GHnbound}, \Cref{TaylorG} and \Cref{TaylorH} we obtain 
	\begin{align*}
		\left\Vert \exp \left( \sum_{j=1}^{2p+1} A_j\right)-h_{n}(\{A_j\})\right \Vert	
		\leq
		\dfrac{1}{3n^2}\left( \sum_{j=1}^{2p+1} \Vert A_j\Vert\right)^3\exp\left(\sum_{j=1}^{2p+1} \Vert A_j\Vert\right).
	\end{align*}
	
\end{proof}

\begin{rmk}
Let $A_1=B_1, A_2=A_3=B_2/2,\cdots, A_{2p}=A_{2p+1}=B_p/2.$	Then {\rm \Cref{thm:suzuki:nonsy}} is reduced to the well-known Suzuki symmetrized approximations for the exponentiated sums 
in {\rm \cite[Formula 3, Equation (1.15)]{suzuki:1985:transfer}}.
\end{rmk}


The following collorary is another generalized Lie-Trotter formula for an 
arbitrary number of elements in Banach algebras. 

\begin{cor}
	For any finite number of elements $A_1, A_2, A_3\cdots, A_{2p}, A_{2p+1}$ in a Banach algebra $\Al,$	
	\begin{align*}
		\lim\limits_{n\to \infty}h_{n}\left(\{A_j\}\right)=\exp\left(\sum_{j=1}^{2p+1} A_j\right).	
	\end{align*}
\end{cor}


\end{document}